\documentclass[aps,pre,onecolumn,notitlepage]{revtex4-1}
\usepackage{epsfig,graphics,amssymb,amsmath,subeqnarray,color}

\def \bsigma{\boldsymbol{\sigma}}

\def \hb{\bar{h}}

\def \UD{U_\Delta}

\def \bF{\mathbf{F}}

\def \bn{\mathbf{n}}

\def \la{\left<}
\def \ra{\right>}
\def \ei{\eta_1}
\def \eii{\eta_2}
\def\d{{\rm d}}

\begin{document}

\title{Synchronization of flexible sheets}
\author{Gwynn J. Elfring}
\author{Eric Lauga\footnote{Corresponding author. Email: elauga@ucsd.edu}}
\affiliation{
Department of Mechanical and Aerospace Engineering, 
University of California San Diego,
9500 Gilman Drive, La Jolla CA 92093-0411, USA.}
\date{\today}
\begin{abstract}
When swimming in close proximity, some microorganisms such as spermatozoa  synchronize their flagella. Previous work on swimming sheets showed that such synchronization requires a geometrical asymmetry in the flagellar waveforms. Here we inquire about a physical mechanism responsible for such symmetry-breaking in nature. Using a two-dimensional model, we demonstrate that flexible sheets with symmetric internal forcing, deform when interacting with each other via a thin fluid layer in such a way as to systematically break the overall waveform symmetry, thereby always evolving to an in-phase conformation where energy dissipation is minimized. This dynamics is shown to be mathematically equivalent to that obtained for prescribed waveforms in viscoelastic fluids, emphasizing the crucial role of elasticity in symmetry-breaking and synchronization.
\end{abstract}
\maketitle

\section{Introduction}
Motile microorganisms swim in a fluid regime where inertia is unimportant and viscous stresses dominate. In this limit the flow field due to a swimmer affects the motility of nearby cells \cite[]{lauga09b}, a fact which is biologically important as  microorganisms such as spermatozoa are often found in high-density suspensions \cite[]{suarez06}. A particular consequence of these fluid-based interactions is the synchronization of the flagella of some spermatozoa observed to occur when these cells are swimming in close proximity \cite[]{woolley09,hayashi98,riedel05}.

G.I. Taylor provided the first quantitative analysis of this phenomenon in his landmark work on the mechanics of swimming microorganisms \cite[]{taylor51}. Using a two-dimensional model of sheets passing sinusoidal waves of transverse displacement, he showed that the energy dissipated by the fluid due to such motions, for two swimmers a fixed distance apart, is minimized if they are in-phase \cite[]{taylor51}. Subsequent computational studies \cite[]{yang08,fauci90,fauci95} have shown that such synchronization can occur due to fluid forces alone.

In recent theoretical analysis it was demonstrated that two infinite sheets passing waves of a prescribed shape, will not synchronize in a Newtonian fluid if the shape of the waveforms $\ei$ and $\eii$ satisfy $\eii(x)=-\ei(-x+\theta)$ (where $\theta$ is a fixed phase shift and $x$ is the direction along the sheets) because of the kinematic reversibility of the Stokes equations \cite[]{elfring09,elfring10b}. The sinusoidal waveforms of Taylor's swimming sheet clearly fall into this category, and will thus not dynamically synchronize in a Newtonian fluid. It has been observed that excess symmetry similarly curbs synchronization in other models \cite[]{kim04,pooley07,putz09, uchida11, golestanian11}. For a sinusoidal sheet, a geometric perturbation must therefore  be added (for example in the form of a higher order mode) to break the necessary front/back symmetry, and  give rise to a time-evolution of phase toward the synchronized state \cite[]{elfring09,elfring10b}. Alternatively, instead of a geometric symmetry breaking, it has also been shown that synchronization can occur if the kinematic reversibility of the field equations is removed, as is the case for  a viscoelastic fluid \cite[]{elfring10}. In such a scenario the phase always evolves to a stable in-phase conformation where the energy dissipated by the swimmers is minimized.

In this paper we inquire about a physical mechanism responsible for symmetry-breaking in real biological cells. Instead of delineating a fixed waveform for the swimming sheets, we take the more realistic modeling approach of passing internal waves of bending as produced by a flagellum's internal structure (or axoneme) \cite[]{riedel07}. In the case of a single sheet, similar models  have been employed to study swimming \cite[]{argentina07,balmforth10} and peeling \cite[]{hosoi04}. We use this model to show that elastic deformation due to fluid body interactions, with purely sinusoidal forcing, always leads to in-phase synchronization.
 
Flexibility has long been considered as an avenue for symmetry breaking in Stokes flow. Purcell, in his celebrated talk and paper \cite[]{purcell77}, asserted that while a stiff oar undergoing reciprocal motion would produce no net motion, due to the scallop theorem, a flexible oar would escape this conundrum because of a broken symmetry between forward and reverse strokes. This was first investigated analytically by Machin \cite[]{machin58} while more recently a number of theoretical and experimental studies have further elucidated the effect of flexibility quantitively, both for boundary-driven \cite[]{wiggins98,yu06,lauga07} and internally-driven filaments  \cite[]{camalet99,camalet00,dreyfus05}.

Recently, the analysis of a pair rotating helices as model for bacterial flagella has shown flexibility to be a crucial ingredient for synchronization \cite[]{reichert05}, as was similarly shown for a minimal model of interacting cilia \cite[]{niedermayer08}. Additionally flexibility has been found to be requisite for the synchronization of paddles that would otherwise be too symmetric to yield stable fixed points \cite[]{qian09}. In these models and experiments, the bodies are rigid, but permitted to deviate from their trajectories, in an elastic manner in response to fluid forces. In contrast, here we allow the bodies themselves to deform due to fluid stresses induced by the other swimmer. 

Our approach is organized as follows. For a pair of two-dimensional sheets in the lubrication limit, we derive a system of nonlinear equations governing both the fluid stresses and the resulting swimmer shapes. We linearize these equations to produce analytical solutions, and then solve the nonlinear equations numerically. We show that flexible sheets with symmetric sinusoidal forcing will deform when interacting with each other via a thin fluid layer in such a way as to break geometrical symmetry, and to evolve to an in-phase conformation where energy dissipation is minimized. Further, this evolution of phase is shown to be functionally equivalent to that found for prescribed waveforms in viscoelastic fluids \cite[]{elfring10}, illuminating the role of elasticity in symmetry-breaking and synchronization.

\section{Model system}
\begin{figure}
\centerline{\includegraphics[width=0.65\textwidth]{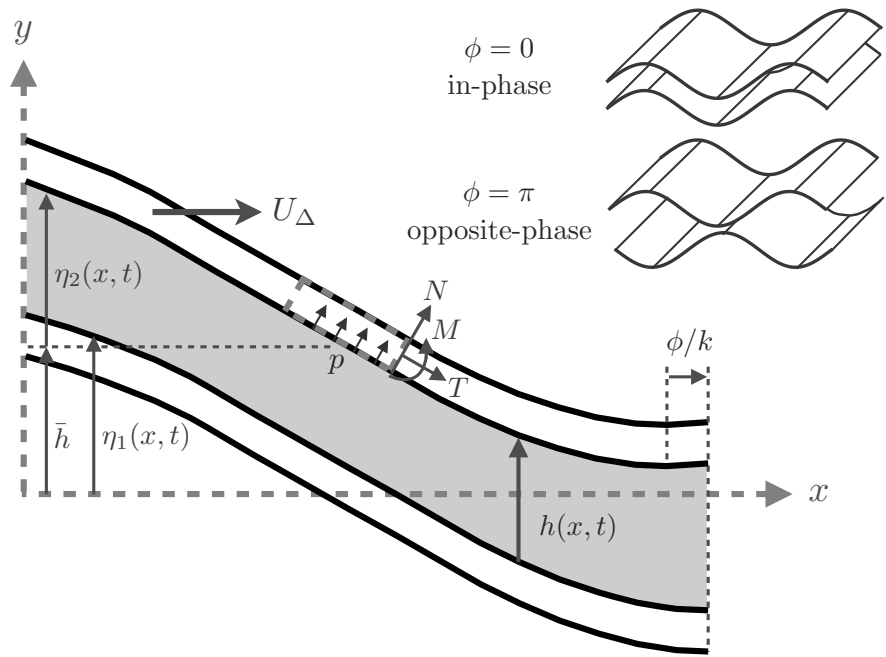}}
\caption{System of two infinite two-dimensional sheets, of shape described by the functions $\ei$ and $\eii$, which  are separated by a fluid layer of thickness $h$ (mean value, $\hb$), have a relative phase $\phi$,  and may move relative to each other with a velocity $\UD$. An infinitesimal material element on the top sheet is subject to fluid pressure $p$, normal force $N$, tension $T$ and moment $M$. Inset: schematic representation of in-phase and opposite-phase configurations. }
\label{system}
\end{figure}

We consider the dynamics of two infinite two-dimensional elastic sheets which are separated by a fluid layer of mean distance $\hb$ (see figure \ref{system}). The sheets deform due to a balance between an active moment $m$, passive bending (elastic) resistance, and fluid stresses. The positions of the sheets are given by $y_1=\ei(x,t)$ for the bottom sheet and $y_2=\hb+\eii(x,t)$ for the top sheet. We look to solve this problem in the limit that the fluid layer is thin compared to the wavelength of the sheets, $k\hb\ll 1$, to make use of Reynolds' lubrication approximation for the fluid field equations \cite[]{reynolds86}.

An infinitesimal element along one sheet is subject to tension, $T$, normal force, $N$ and moment, $M$ (shown in figure \ref{system}). Since the fluid layer is thin and lubrication forces are singular with the gap thickness, forces from the outer flow are safely ignored \cite[]{balmforth10}. To capture the active bending of the sheet we use a model for the flagella of eukaryotes  introduced by J\"ulicher and co-workers \cite[]{camalet00,riedel07} where the bending of an elastic filament (flagellum) is caused by the constrained sliding of microtubule doublets; the effect of the internal forces which induce sliding are here represented by an active moment density $m$. Given that we are in an over-damped limit we take the equilibrium shapes to arise instantaneously \cite[]{hosoi04}. Assuming reasonably small deflections such that the swimmers are linearly elastic \citep[]{landau86}, force ($\bF$) and moment (M) balance on the top sheet yield respectively
\begin{eqnarray}
\frac{\partial \bF}{\partial x}=-\bn\cdot\bsigma,\quad\quad \frac{\partial M}{\partial x}=-N+m,
\label{balances}
\end{eqnarray}
where $\bn$ is the unit normal and $\bsigma$ is the fluid stress tensor. The relation between the moment and the sheet curvature is given by the constitutive relation $M=B\partial^2 \eta /\partial x^2$, where $B$ is the sheet bending stiffness. Combining with \eqref{balances} we obtain the equations governing the shape of the sheets
\begin{eqnarray}
B\frac{\partial^4 \eta}{\partial x^4}=\frac{\partial m}{\partial x}\mp \bn\cdot\bsigma\cdot\bn,
\label{beam1}
\end{eqnarray}
where $\mp$ are for the top and bottom sheets respectively. If resistive force theory is used for the fluid forces then we obtain the governing equation used by \cite{riedel07} for a single filament.

The internal forcing on the sheets is assumed to take the form $m(x,t)=A g(kx-kct)$ where $k$ is the wavenumber, $c$ is the wave speed, $A$ is the amplitude of the moment and $g$ is an arbitrary but $2\pi$-periodic function. Because the forcing is periodic, we will assume the  shape $\eta$ to also  be periodic. We nondimensionalize vertical distances by $y^*=y/\hb$ and horizontal distances $x^*=kx$ ($^*$ indicates a dimensionless quantity). Nondimensionalizing the continuity equation we find that if the horizontal velocity is given by $u=cu^*$ then the vertical velocity must be $v=\epsilon c v^*$ where $\epsilon=k\hb$. The Stokes equations then yield the lubrication equations to leading order in $\epsilon$
\begin{gather}
\frac{\partial p^*}{\partial x^*}=\frac{\partial^2 u^*}{\partial y^{*2}}, \quad\quad \frac{\partial p^*}{\partial y^*}=0,\\
\frac{\partial u^*}{\partial x^*}+\frac{\partial v^*}{\partial y^*}=0.
\label{lubrication}
\end{gather}
where $p^*=\epsilon^2p/ \mu\omega$. Forces (per unit depth) are nondimensionalized as $f^*= f\epsilon/\mu c$, while energy dissipation rate per unit depth is $\dot{E}^*=\epsilon^2\dot{E}/\mu\omega c\hb$.

In the lubrication limit, $\epsilon \ll 1$, the normal force due to the fluid on the beam is to leading order merely the pressure, $-\bn\cdot\bsigma\cdot\bn=p$. Since the field equations for the fluid yield the pressure gradient, we differentiate \eqref{beam1}, and recasting the equation in dimensionless form we obtain
\begin{eqnarray}
B^*\frac{\partial^5\eta^*}{\partial x^{*5}}=A^*\frac{\partial^2 g}{\partial x^{*2}}\pm\frac{dp^*}{dx^*},
\label{beam2}
\end{eqnarray}
where $B^*=B\epsilon^3k^3/\mu\omega$ is the dimensionless bending stiffness and $A^*=A\epsilon^2k^2/\mu\omega$ is the dimensionless amplitude of the active bending moment. This equation allows us to solve for the shape of the sheets, $\eta_{1,2}$, and is coupled to the fluid field equations through the pressure gradient. We now drop the $^*$ for convenience.

\section{Analysis}
Given the form of the forcing we expect post-transient solutions which are functions of a wave variable $z=x-t$ and thus we write $\eta=\eta(z)$. The top sheet may move relative to the bottom sheet with a horizontal velocity $u=\UD$, hence the boundary conditions for the fluid equations \eqref{lubrication}, in a frame moving with waveform are given by $u(x,y_1)=-1$, $v(x,y_1)=-\ei'$, $u(x,y_2)=U_\Delta-1$ and $v(x,y_2)=-\eii'$. Given the above boundary conditions the solution for the velocity field is found to be
\begin{eqnarray}
u(x,y)=\frac{1}{2}\frac{dp}{dx}(y-y_1)(y-y_2)+\UD\frac{y-y_1}{y_2-y_1}-1.
\label{u}
\end{eqnarray}
If one integrates the continuity equation one finds
\begin{eqnarray}
\frac{\partial}{\partial x}\int_{y_1}^{y_2}udy=\UD\frac{d\eii}{dx}\cdot
\label{flowrate}
\end{eqnarray}
If $\UD=0$ then the flow rate between the sheets is constant. Integrating \eqref{flowrate} and exploiting the periodicity of the pressure  \cite[]{elfring09}, we obtain the equation for the pressure gradient
\begin{align}
\frac{dp}{dx}=\frac{6\UD-12}{h^2}-\frac{12\UD y_2}{h^3}-\frac{(6\UD-12)I_2-12\UD J_3}{I_3h^3},
\label{pressure}
\end{align}
where the distance between the two sheets is given $h=1+\eta_2-\eta_1$ and $I_j=\int_{0}^{2\pi}h^{-j}dx$ and $J_3=\int_{0}^{2\pi}y_2 h^{-3}dx$. Then the force on the top sheet is given by
\begin{eqnarray}
f_x=\int_{0}^{2\pi}\left.\left(y_2\frac{dp}{dx}-\frac{\partial u}{\partial y}\right)\right|_{y=y_2}dx
=\int_0^{2\pi} \left(\frac{1}{2}\frac{dp}{dx}(\eii+\ei)-\frac{\UD}{1+\eii-\ei}\right)dx
\label{force}.
\end{eqnarray}

\subsection{Linear regime: Statics}
We first look at the case where we enforce $\UD=0$ (i.e. we fix the top sheet with respect to the bottom sheet) in order to determine under which condition  a nonzero synchronization force will arise. 

If we assume the dimensionless amplitude of the forcing $A$ to be  small, and the dimensionless  bending stiffness $B$ to be large, then the shape amplitude (or maximum value of the shape) $\| \eta \|_\infty$ is expected to also be small. Linearizing the pressure gradient for small $\eta$, with $\UD=0$ gives
\begin{eqnarray}
\frac{dp}{dx}\approx -12(\eta_2-\eta_1),
\label{linpressure}
\end{eqnarray}
where we have invoked an integrated conservation of mass,  $\la \eii-\ei \ra=0$  (angle brackets $\la \ra$, denote the average over a period). Our goal is now to determine whether two sheets which are equally and symmetrically forced but with a phase shift $\phi$ will synchronize in time. With this in mind we let $g_2=g_1(z+\phi)$ with $g_1(z)=\cos(z)$, and we set equal for both sheets the bending stiffness $B$ and the forcing amplitude $A$. The linearized governing equations are then given by
\begin{subeqnarray}
B\frac{d^5 \ei}{d z^5}-12(\eta_2-\eta_1)&=& -A\cos(z),\\
B\frac{d^5 \eii}{d z^5}+12(\eta_2-\eta_1) &=& -A\cos(z+\phi).
\label{linear}
\end{subeqnarray}
To solve these equations we apply periodic boundary conditions. 
The solution of this system of equations with the linearized pressure gradient \eqref{linpressure}, can be found analytically to be
\begin{subeqnarray}
\ei(z)&=&A\frac{12B \big[\cos(z+\phi)-\cos z\big]
-\big[\left(288+B^2\right) \sin z+288 \sin(z+\phi)\big]}{B \left(576+B^2\right)}+C,\quad\\
\eii(z)&=&A\frac{12B \left[\cos z-\cos(z+\phi)\right]
-\left[288 \sin z+\left(288+B^2\right) \sin (z+\phi)\right]}{B \left(576+B^2\right)}+C.
\label{eq:shapes}
\end{subeqnarray}
Note that nothing prevents the solution from including a uniform shift $C(A,B)$; however, the relevant physics of the problem are invariant under such a shift hence and hence the value of  $C$ is irrelevant (equivalently, we place our $z$-axis at $\la \eta_1 \ra=0$). 

Both shapes in \eqref{eq:shapes} are delineated by the competition between bending rigidity, the pressure gradient in the fluid, and the internal forcing (with $\phi$ dependence). In this linear limit we note that the sheets are linear in the forcing amplitude $A$, and when $A=0$ then as expected they become straight, {\it i.e.}~$\eta=0$. In the limit where rigidity dominates, $B\rightarrow \infty$, then the sheets also tend to become straight,  $\eta\rightarrow 0$. If the rigidity and forcing amplitude are both very large ($A,B\gg 1$) then we can scale out the contribution from the fluid forces in \eqref{linear} and we are left with $\ei\approx -(A/B)\sin z$ and $\eii\approx -(A/B)\sin (z+\phi)$ as might be expected. Note finally that the solutions in \eqref{eq:shapes} are only valid when $B\gg A$ as otherwise unphysical solutions may arise with the sheets overlapping; this is  prevented when the full nonlinear form of the pressure gradient is kept, as it diverges when $h\rightarrow 0$.  

\begin{figure}
\centerline{\includegraphics[width=0.75\textwidth]{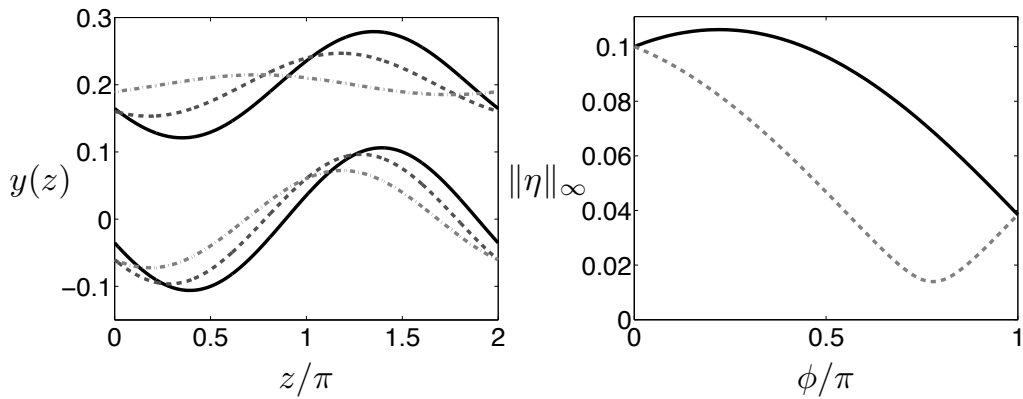}}
\caption{Left: Solution shapes, \eqref{eq:shapes}, for various phase differences $\phi=\pi/4$ (solid lines), $\pi/2$ (dashed lines), and $3\pi/4$ (dash-dot lines), with $A=1$ and $B=10$. We observe that the amplitude is not evenly affected by the pressure (the plot is shown here with $\hb=0.2$ rather than $\hb=1$ for display purposes only). Right: Shape amplitude, $\| \eta \|_\infty$, vs.~phase difference, $\phi$ (bottom sheet: solid line;  top sheet:  dashed line). Lines are reversed upon the change $\phi\rightarrow-\phi$.}
\label{shapes}
\end{figure}

Solutions to \eqref{eq:shapes} for $A=1$, $B=10$, and three values of the phase difference ($\phi=\pi/4, \pi/2, 3\pi/4 $) are plotted in figure \ref{shapes} (left). We observe that the shapes are sinusoidal and hence individually remain symmetric both about the vertical axis and the horizontal axis. The global asymmetry that arises however is that the amplitudes of the two waveforms are not equally modulated by the fluid pressure, as shown in figure \ref{shapes}  (right). We see indeed that the top sheet has smaller amplitude for positive $\phi$ (and by symmetry, this is reversed upon changing $\phi\rightarrow-\phi$).

The phase locking force on the top sheet is, at leading order, given by
\begin{eqnarray}
f_x=-6\int_{0}^{2\pi}(\eii-\ei)(\eii+\ei)dx
=2\pi\alpha\sin\phi.
\label{linforce}
\end{eqnarray}
where $\alpha=144 A^2/(576 B+B^3)$. Equation \eqref{linforce} is the main result of our paper.  
The phase locking force is proportional to the sine of the phase, meaning that the only stable fixed point is expected to occur at $\phi=0$, and hence all initial conformations will evolve to the stable in-phase conformation. We thus see that the elasticity of the swimmers, and thus fluid-structure interactions, can introduce the geometrical symmetry-breaking necessary to develop a nonzero phase locking force. The force is found to be quadratic in amplitude, reminiscent of viscoelastic symmetry-breaking \cite[]{elfring10}; by comparison, the phase-locking force arises at fourth order in amplitude for prescribed asymmetric waveforms in a Newtonian fluid. The reason for the difference is that with elastic deformation, the sheets are ultimately not the same shape, despite having identical mechanical properties, and hence $\la\ei^2\ra\ne\la\eii^2\ra$; for prescribed waveforms this is different as the same waveform is prescribed for both sheets, and the quadratic term of the force vanishes.

The energy dissipated by the fluid between the two swimming cells to leading order is
\begin{eqnarray}
\dot{E}=12\int_{0}^{2\pi}\left(\eii-\ei\right)^2dx
=\frac{24\pi A^2}{576+B^2}(1-\cos\phi).
\end{eqnarray}
We see that the energy dissipation is a global minimum when $\phi=0$ and global maximum when $\phi=\pi$. It follows then that the cells will always evolve to a state of minimum energy dissipation. We observe that the form of the energy dissipation is precisely the same as that for  fixed shapes (and taking the fixed wave amplitude $A_{fixed}^2=A^2/(576+B^2)$ they are equal) \cite[]{elfring10b}. It is important to note that waveforms with a prescribed broken symmetry may evolve to either the in-phase or opposite-phase conformation \cite[]{elfring09}; in contrast, the natural symmetry-breaking due to elasticity of the bodies, or in the fluid, leads to a conformation of minimum energy dissipation.

\subsection{Linear regime: Dynamics}
When the sheets are permitted to evolve in time in force-free swimming, the relative velocity $\UD$ will thus be nonzero. In order to determine the leading order component of the pressure field we must first find out how the relative velocity scales with the sheet amplitudes. Given that the net force on the sheets in the dynamic case must now be zero, we obtain at leading order the relative speed as given by
\begin{eqnarray}
\UD=-\frac{3}{\pi}\int_{0}^{2\pi}(\eii-\ei)(\eii+\ei)dx.
\label{linvelocity}
\end{eqnarray}
We see that the velocity is quadratic in amplitude and indeed is proportional to the static force $\UD=f^s_x/2\pi$ (we use here the superscript \textit{s} to indicate the static force given by \eqref{linforce} to avoid confusion). With this we thus know that at leading order the pressure field (for a given $\phi$) is invariant between the static and dynamic case, and since the beam equation couples via the pressure field, our instantaneous shapes are found to be the same. 
The only difference is that now the phase difference changes in time (geometrically) due to the presence of a nonzero relative velocity, according to $\d \phi/ \d t=-\UD$. Using \eqref{linvelocity}, we find the rate of change of the phase at leading order to be given by
\begin{eqnarray}\label{diff}
\frac{d\phi}{dt}=-\alpha\sin\phi.
\label{phasede}
\end{eqnarray}
Equation \eqref{diff} can be integrated analytically, leading to a time-evolution of the phase given by
\begin{eqnarray}
\phi(t)=2\tan^{-1}\left[\tan\left(\frac{\phi_0}{2}\right)e^{-\alpha t}\right].
\label{phasetime}
\end{eqnarray}
All initial conformations, $\phi_0$, decay in time to the stable in-phase conformation, $\phi=0$. Notably,  the time-evolution of the phase for a sinusoidally forced elastic sheet we obtain here is mathematically similar to that for a fixed sinusoidal waveform in a viscoelastic fluid \cite[]{elfring10} and for rigid bodies with flexible trajectories \cite[]{niedermayer08}, emphasizing therefore the crucial role of elasticity in synchronization.

Finally, if we allow a small difference in the wavespeeds of the sheets, $\Delta\omega$, then to leading order we have the same evolution of phase, \eqref{phasede} and \eqref{phasetime}, but now the rate of change of phase is defined as $\dot{\phi}=-\UD-\Delta\omega$ and hence we see a synchronization of shape but not of material points.
\subsection{Nonlinear Case}
\begin{figure}
\centerline{\includegraphics[width=.75\textwidth]{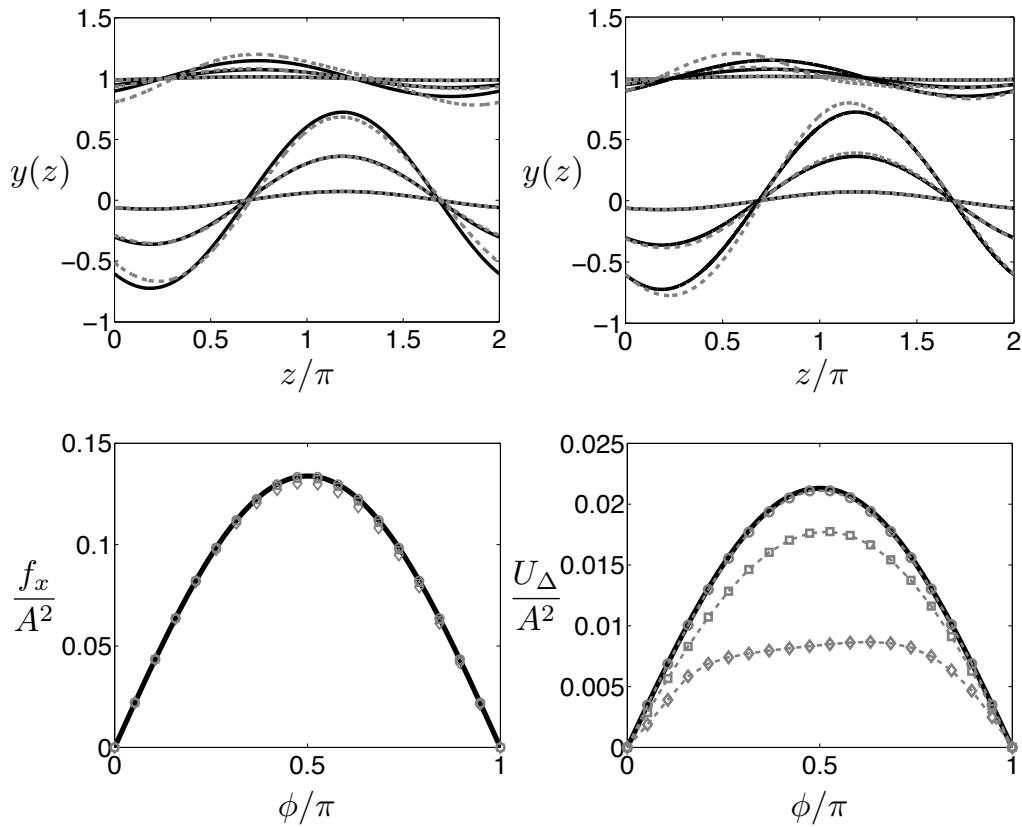}}
\caption{Top: solution shapes, $\eta_1$ and $\eta_2$, for bending stiffness $B=10$, phase difference $\phi=3\pi/4$, and amplitudes $A=\{1,5,10\}$ (left: $\UD=0$;  right $f_x=0$);  linearized pressure approximation: solid lines;  full pressure gradient: dashed  lines. Bottom left:  phase locking force, $f_x/A^2$, vs.~phase difference, $\phi$, when $\UD=0$;  bottom right:  relative velocity, $\UD/A^2$, vs.~$\phi$, when $f_x=0$. Both plots are for numerical solutions of the nonlinear equations with $B=10$ and $A=1$ (circles), 5 (squares), and 10 (diamonds);  linearized solutions are shown as solid lines. Away from the linear  regime, the rate of change of the phase is affected while the forces are not.}
\label{shapeslarge}
\end{figure}

To move beyond the linear regime, we now solve the nonlinear equations for the shapes $\eqref{beam2}$ numerically, together with  \eqref{pressure}, using Matlab's  boundary value problem solver \texttt{bvp4c}, both for the static case, $\UD=0$, and the force free case, $f_x=0$ using \eqref{force}.

We find the linearized pressure gradient to be a capable approximation, particularly when the bending is of the same order as the pressure $B\sim1$ and the phase difference is small; however, when $A,B\gg1$ the linearized pressure may lead to unphysical solutions particularly if the sheets are near opposite-phase as the divergent nature of the full form of the pressure gradient is required to deform the sheets from contact. 

In figure \ref{shapeslarge} we illustrate  the breakdown of the linear regime. We plot the static shapes ($\UD=0$, top left), and dynamic shapes ($f_x=0$,  top right), both with a phase difference of $\phi=3\pi/4$, bending stiffness $B=10$, and with forcing amplitudes $A=\{1,5,10\}$. We see that for increasing amplitude the shapes predicted by linearized equations (solid) and nonlinear equations (dashed) begin to diverge. In particular the nonlinear equations lead to a pronounced left-right asymmetry in the individual shape. 
In the lower left plot of figure \ref{shapeslarge} we display the phase locking force vs.~phase, while in the lower right plot we show the relative velocity vs.~phase, both with $B=10$ and $A=\{1,5,10\}$ (circles, squares and diamonds respectively) for the numerical solutions to the nonlinear equations; the analytical solutions for the linear equations are shown solid. The synchronizing hydrodynamic force, $f_x$, found by either method remains remarkably consistent even for very large forcing amplitude, $A$. In contrast, the rate of change of the phase decreases markedly from the linear approximation for large amplitude waves. Because the force is virtually unaffected we know therefore that the resistance to motion is dramatically increased by the change in shape.

\begin{figure}
\centerline{\includegraphics[width=0.75\textwidth]{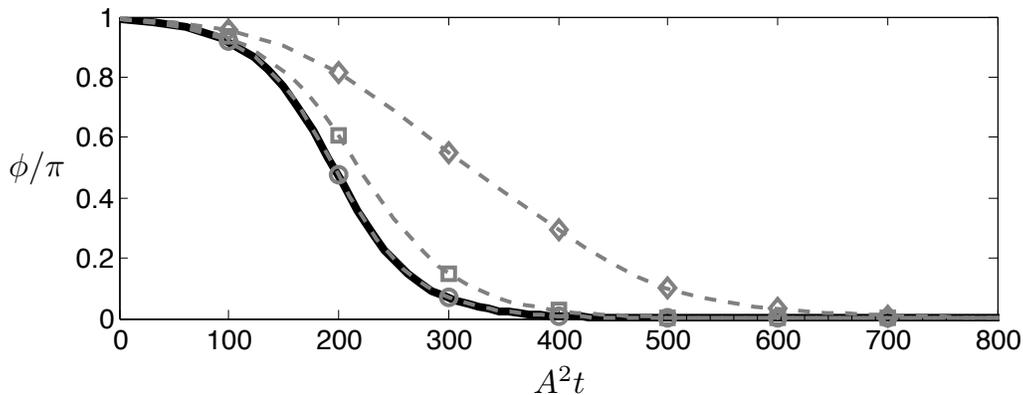}}
\caption{Time-evolution of the phase difference, $\phi$,  in the  nonlinear  problem for $B=10$ starting from an initial angle of $\phi/\pi=.99$ and with forcing amplitude $A=1$ (circles), 5 (squares) and 10 (diamonds); the linear estimate is shown solid. As the forcing amplitude increases, the linear solution increasingly underestimates the time scale to synchronize.}
\label{dynamics}
\end{figure}

In figure \ref{dynamics} we integrate the instantaneous relative velocity to obtain the evolution of the phase in time. We show solution to both the linear equations (solid) and nonlinear equations (dashed) for $B=10$ and $A=\{1,5,10\}$ starting from an initial phase difference of just less than $\pi$. We see that, as the forcing amplitude increases, the nonlinear equations yield an increasingly slower evolution to a synchronized conformation than that predicted by the linear regime; however, the general behavior remains qualitatively unchanged.

\section{Conclusion}

In this paper we inquired about a physical mechanism responsible for symmetry-breaking and synchronization in the flagella of biological cells such as spermatozoa. In a Newtonian fluid, two swimming sheets passing waveforms of a prescribed sinusoidal shape will not synchronize due to an excess of symmetry; however, here we have demonstrated that identical flexible sheets with symmetric sinusoidal forcing will deform, when interacting with each other via a thin fluid layer, in such a way as to systematically break the overall geometrical symmetry. This system will always evolve to an in-phase conformation where energy dissipation is minimized, in contrast to a prescribed asymmetry, which may maximize energy dissipation. In addition, this time-evolution of the relative phase is shown to be equivalent to that obtained for prescribed waveforms in viscoelastic fluids \cite[]{elfring10}, emphasizing the crucial role of elasticity in symmetry-breaking and synchronization --  be it that of the fluid, or the swimmers themselves.

\begin{acknowledgments}
Funding by the NSF (CBET-0746285) and NSERC (PGS D3-374202) is gratefully acknowledged.
\end{acknowledgments}

\bibliography{phaselocking}

\end{document}